\documentclass{emulateapj}   

\begin{document}

\title{The Santa Fe Light Cone Simulation Project: II. The Prospects for Direct Detection of the WHIM with SZE Surveys}
\author{Eric J. Hallman\altaffilmark{1,2}, Brian W. O'Shea\altaffilmark{3}, 
Britton D. Smith\altaffilmark{2}, Jack O. Burns\altaffilmark{2} \& Michael L. Norman\altaffilmark{4}}

\altaffiltext{1}{National Science Foundation Astronomy and
  Astrophysics Postdoctoral Fellow}

\altaffiltext{2}{Center for Astrophysics and Space Astronomy,
  Department of Astrophysics and Planetary Sciences, 
  University of Colorado at Boulder, Boulder, CO 80309; hallman, burns@casa.colorado.edu}

\altaffiltext{3}{Department of Physics and Astronomy, Michigan State
  University, East Lansing, MI; oshea@msu.edu}

\altaffiltext{4}{Center for Astrophysics and Space Sciences,
University of California at San Diego, La Jolla, CA 
92093; mnorman@cosmos.ucsd.edu}

\begin{abstract}
Detection of the Warm-Hot Intergalactic Medium (WHIM) using
Sunyaev-Zeldovich effect (SZE) surveys is an intriguing possibility,
and one that may allow observers to quantify the amount of ``missing
baryons'' in the WHIM phase. We estimate the necessary sensitivity for
detecting low density WHIM gas with the South Pole Telescope (SPT) and
Planck Surveyor for a synthetic 100 square degree sky survey. This
survey is generated from a very large, high dynamic range adaptive
mesh refinement cosmological simulation performed with the Enzo
code. We find that for a modest increase in the SPT survey sensitivity
(a factor of 2-4), the WHIM gas makes a detectable contribution to the
integrated sky signal. For a Planck-like satellite, similar detections
are possible with a more significant increase in
sensitivity (a factor of 8-10). We point out that for the WHIM gas, the
kinematic SZE signal can sometimes dominate the thermal SZE where the
thermal SZE decrement is maximal (150 GHz), and that using the
combination of the two increases the chance of WHIM detection using
SZE surveys. However, we find no evidence of unique
features in the thermal SZE angular power spectrum that may aid in its
detection. Interestingly, there are differences in the power spectrum
of the kinematic SZE, which may not allow us to detect the WHIM
directly, but could be an important contaminant in cosmological
analyses of the kSZE-derived velocity field.  Corrections derived from
numerical simulations may be necessary to account for this
contamination. 
\end{abstract}


\section{Introduction}\label{sec:Intro}
A significant fraction of baryons in the universe are expected to be
in the Warm-Hot Intergalactic Medium (WHIM). Numerical cosmological
simulations \citep[e.g.,][]{dave, cen} predict that somewhere around
30-40\% of all baryons exist in this phase, characterized by electron
temperatures between $10^5$ and $10^7$K. This shock-heated gas primarily is located
in large-scale filaments and sheets in the cosmic web, strung between
clusters of galaxies. Observations of this gas (if collisional
ionization is the correct interpretation) have been
made in Ly$\alpha$ and OVI absorption \citep{tripp, danforth05,
  danforth08}, and also are likely detectable in UV absorption lines
and X-ray lines of OVII and OVIII \citep{nicastro}. Detection of these
so-called ``missing baryons'' is critical to understanding the matter
content of the universe.   

Recently it has been suggested that the WHIM gas may be detectable
using the Sunyaev-Zeldovich Effect (SZE) \citep{monteagudo, sflc,
  afshordi}. The SZE is a consequence of cosmic microwave background
(CMB) photons being inverse Compton scattered to higher energy by hot
electrons. This upscattering creates a low frequency decrement in the
CMB, and a corresponding higher frequency increment \citep[for review,
see e.g.,][]{reph,birk, carlstrom}. The SZE is most typically discussed as it relates to
observations of clusters of galaxies, since the very hot ($\sim
10^8K$) gas there is an effective source of high velocity scattering
electrons. Detections and analysis of known galaxy clusters have been
made by SZE 
telescopes \citep[e.g.,][]{laroq,apex_bull} and a recent blind search
using the South Pole Telescope (SPT)
has detected previously unknown galaxy clusters \citep{spt_det}. Investigators have attempted to estimate the SZE signature
of galaxy groups \citep{moodley} and have made weak detections of
supercluster gas in Corona Borealis \citep{battis,genova}. In
addition, much work has been done to study the possibility of using
the kinematic SZE to determine both cluster peculiar velocities
\citep{repha_k,haeh,agha} and the overall bulk velocity field \citep[e.g.,][]{kash}.

 Recent work suggests that filaments in large-scale
structure, because of their long path lengths and correlated
velocities, may have a unique angular power signature in the SZE
\citep{atrio}. In this paper, we show that the SZE may contribute a
detectable signal from the WHIM gas, observable with
extensions to the sensitivity of current and upcoming surveys (e.g.,
SPT, Planck Surveyor) over relatively small angular areas.  We examine the contribution of the
filamentary WHIM to the angular power spectrum of both the kinematic and
the thermal SZE.
\subsection{The Relative Strengths of Thermal and Kinematic SZE}
The SZE can be separated into two spectrally distinct
effects, one which results from the scattering of the CMB by the
electrons at the thermal velocity in the gas, and the other resulting
from the bulk line-of-sight velocity of the gas (kinematic
SZE). Typically for the purposes of clusters of galaxies, the
kinematic SZE (kSZE) can
be considered a small perturbation on the full SZE signal. However, in the
cooler WHIM gas the relative strength of the kSZE increases, becoming
comparable to the thermal SZE (tSZE). The reason for this is the different
dependency of the kSZE and tSZE on the physical properties of the
gas. Indeed, the effects should be comparable when the bulk radial
velocity of the gas is roughly equal to the average thermal velocity of the
electrons in the gas. The thermal SZE signal, characterized by the
Compton y parameter 
\begin{equation}
y = \frac{\sigma_Tk_B}{m_ec^2}\int n_e T_e dl
\end{equation}
depends linearly on $n_e$ and $T_e$ (i.e., is proportional to the
thermal pressure), while the kinematic SZE,
characterized by
\begin{equation}
b = \frac{\sigma_T}{c} \int v_r n_e dl
\end{equation}
depends linearly on electron density and the line-of-sight peculiar (radial) velocity of the
gas. For the kSZE, a negative radial velocity (toward the observer)
results in a positive $\Delta$T/T. For the purposes of this work, we study the SZE signals at 150
GHz (2.1 mm), the frequency where the thermal SZE signal has its
maximum decrement against the CMB. For the kSZE, the relative
temperature shift, $\Delta T/T$, is independent of frequency, and is
equal to the value of $b$. The tSZE value for $\Delta T/T$ is 
\begin{equation}
(\Delta T/T)_{tsze}= y \left( x\frac{e^x+1}{e^x-1} - 4 \right), 
\end{equation}  
which at 150 GHz is equal to $-y$. Here $x = h\nu/kT$. The thermal SZE
is subject to relativistic corrections, which can be important for
very hot ($>10keV$) clusters \citep{itoh}.

If we take the ratio of $b/y$, the relative strength of the SZE
temperature perturbations at 150 GHz, for a given parcel of gas we get
\begin{equation}
\frac{b}{y} = \frac{m_e c}{k_B} \frac{v_r}{T_e}.
\end{equation}
This expression can be written as
\begin{equation}
\frac{b}{y} = 0.197 \frac{v_r/(100km/s)}{T_e/(10^7 K)}.
\end{equation}
A similar calculation is shown in \citet{birk}. For a radial
velocity of 100 km/s, and a gas temperature of $5 \times 10^7$ K (a
temperature typical of clusters), this
ratio is around 4 percent. But in a gas having the same radial velocity
but a reduced temperature, say a
value of T = $10^6$ K, the ratio is roughly 2. Therefore, for million
degree gas with a bulk velocity of 100 km/s along the line of sight,
the magnitude of the kSZE component is twice as large as the tSZE. One
caveat is that kSZE can be either positive or negative, thus could
cancel out the tSZE in some instances. It is possible for the kSZE to
be zero when the gas has no peculiar line-of-sight velocity. In
addition, gas in
cosmological filaments is of relatively low density. 
However, this gas tends to be more spatially extended, thus
there can be very long path lengths,
creating a stronger SZE signal.  We should note that $\Delta T/T$ for
the tSZE is frequency dependent, and on the decrement side at 30 GHz
(where interferometers like CARMA operate), $\Delta T/T$ is twice the
value of $y$, while the kSZE has no frequency dependence.  Thus at
lower frequencies, the ratio of $\Delta T/T$ between kSZE and tSZE is
a factor of 2 lower than the $b/y$ indicated above. 

It is also perhaps important to note that calculations by \citet{fox}
and \citet{yosh05} for example, show that in hot gas, the electron
temperature may have a non-negligible equilibration time with the
(hotter) temperature of the ions in the post-shock regions in filaments. A
systematically low electron temperature impacts the thermal SZE signal
directly, as it is linear with $T_e$. The kSZE is unaffected, as it is
independent of $T_e$. 

\begin{figure}
\includegraphics[width=3.5in]{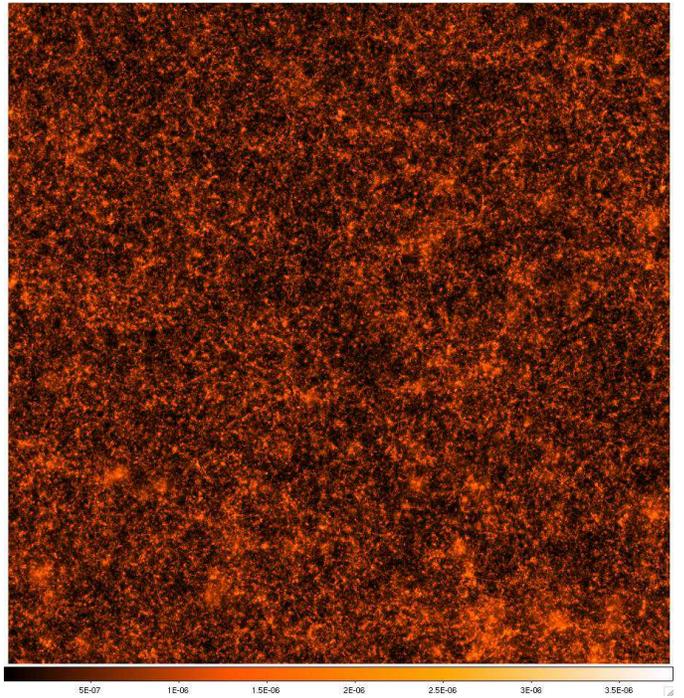}
\caption{Light cone projection for the low-density WHIM (LDW) for the
  thermal SZE. LDW is defined as $10^5 < T < 10^7$K and $0< \delta < 50$. Field spans 10x10 degrees.}
\label{panel1}
\end{figure}
\begin{figure}
\includegraphics[width=3.5in]{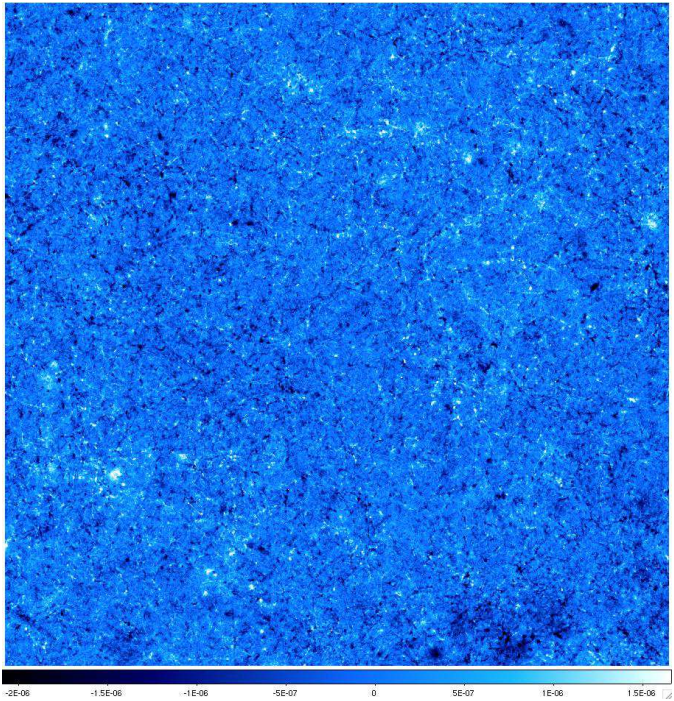}
\caption{Light cone projection for the LDW for the kinematic SZE. Same
field as figure \ref{panel1}}
\label{panel2}
\end{figure}

In any case, it is intriguing that the kSZE may serve to boost the
tSZE signal of the WHIM gas significantly, perhaps aiding in its
detection in a statistical sense through all-sky surveys at the SZE
decrement maximum. While it is sometimes advantageous to 
separate the kSZE and the tSZE by their spectral signatures in order
to measure each effect individually, here we also attempt to
take advantage of their (sometimes) additive effect in order to aid in
the statistical detection of the WHIM gas. 

\section{Simulations}
The main results shown in this work are from analysis of the Santa Fe
Lightcone, described in \citet{sflc}. Briefly, this N-body/hydro
cosmological simulation was carried out with the Enzo \citep{enzo} code for the
express purpose of generating 100 square (10 x 10) degree synthetic
sky surveys in the SZE and X-rays. The calculation is initialized at
$z$=99 using a \citet{eishu99} power spectrum with $n$ = 1. The
cosmological parameters are $\Omega_M = 0.3$, $\Omega_b$ = 0.04,
$\Omega_{CDM} = 0.26$, $\Omega_{\Lambda}= 0.7$, $h=0.7$, and $\sigma_8
= 0.9$. We simulated a cubic volume of dimension 512$h^{-1}$ comoving
Mpc with a root grid of 512$^3$ computational zones and seven levels
of adaptive refinement using the criteria of baryon and dark matter
density. We use 512$^3$ dark matter particles, resulting in a dark
matter mass resolution of $\approx$7.3$\times10^{10} M_{\odot}$.  

The light cones are made by stacking 27 slices of the simulation data,
correcting for their angular scale as a function of their physical
size, and taking random projection axis selections and projection
translations to get multiple realizations of the light cone for the
given cosmology. These slices are projections at various redshifts
spanning the range 3.0 $>$ z $>$ 0. For this work we use 100
realizations of the light cone with random projection axes and
shifts. The final light cone images for this work have image
resolution of 4096$^2$ for an effective resolution of 8.8$\arcsec$
per pixel. From our previous work, we have shown that this simulation
effectively resolves the halo mass function down to roughly
5$\times10^{13} M_{\odot}$. For more details see \citet{sflc}.    

\subsection{Low-Density WHIM Projections}
\begin{figure}
\includegraphics[width=3.39in]{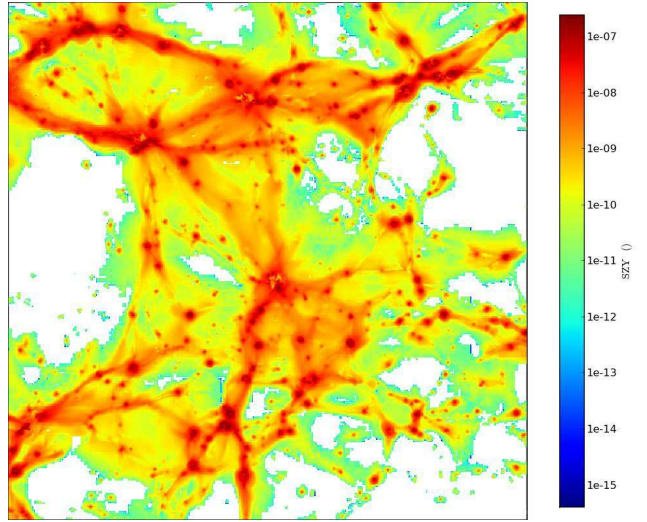}
\caption{Image of the Compton y parameter for the thermal SZE using
  the low-density WHIM (LDW) gas as described in the text. This
  projection is made from a slice of the simulation box at z=0.15,
  with a width subtending 10 degrees at that redshift (comoving
  physical width of $\approx 77h^{-1}$Mpc), and with a depth of
  $\approx 128h^{-1}$Mpc comoving. Illustrates how this range of
  density and temperature captures the filaments.Projection was made using the YT analysis toolkit \cite[\texttt{yt.enzotools.org}]{SciPyProceedings_46}.}
\label{lowz_whim}
\end{figure}

For this particular work, we have used projections of the thermal and
kinematic SZE. We have made light cones
from projections of the kSZE and tSZE from gas within the WHIM
temperature range ($10^5 -10^7K$), and also limit the gas overdensity
to the regime from 0$<\delta<$50 (where $\delta$ indicates the overdensity (gas + dark matter) ratio with respect to $\rho_c$). The limit of 50 serves to
remove gas that may be very near clusters. In a $\Lambda$CDM cosmology with the currently
favored parameter values, this is more than a factor of 3 below the
virial overdensity. Therefore the gas we are probing is definitely in
the filaments and sheets (and even voids). In our tests of overdensity
threshold, we find very weak dependence of the result when changing
the overdensity cutoff by a factor of a few. We hereafter refer to this
gas as the low-density WHIM (LDW). The LDW light cone projections for
the thermal and kinematic SZE are shown in Figures \ref{panel1} and
\ref{panel2}. One reason for the safety
factor in selected overdensity is that in any simulation,
we have a finite mass resolution, which means that low mass halos are
unresolved gravitationally. Therefore the overdensity achieved in the
unresolved halos is likely lower than the typical virial value. The
main purpose for our choice of limits, therefore, is to ensure that we are fully in the
filamentary WHIM regime. Figure \ref{lowz_whim} shows a projection of
a low redshift slice made with the gas properties defined in this way,
illustrating how this regime of temperature and density captures the
filamentary gas.

\section{Detecting the Signature of the WHIM with Planck and SPT}
\begin{figure}
\includegraphics[width=3.0in]{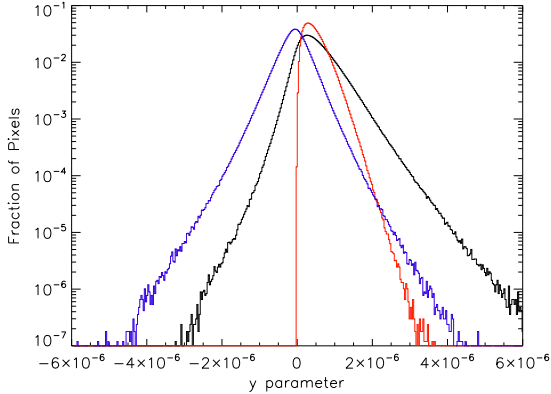}
\caption{Histogram of one of the LDW light cone images in Compton y
  and/or b parameter. Red line indicates the thermal SZE, blue is for
  the kinematic SZE, black is the total of the two, an indicator of
  the total value of $\Delta T/T$ at the thermal decrement maximum
  ($\approx 2.1$mm/150GHz).}
\label{whim_hist}  
\end{figure}

\begin{figure}
\includegraphics[width=3.0in]{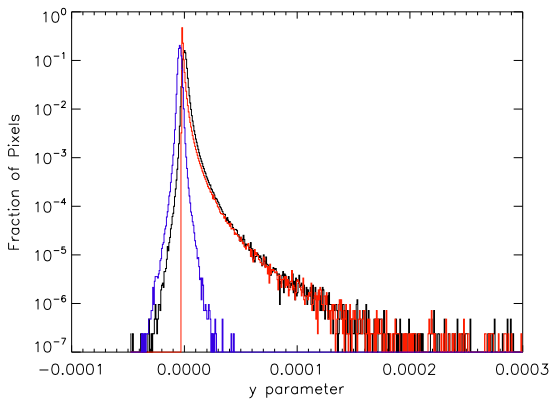}
\caption{Same as Figure \ref{whim_hist}, but for the full light cone
  image including clusters. Red is thermal SZE, black total combined
  signature, and blue is kinematic SZE. Note the relative lack of
  importance of the kSZE at the high decrement end. This tail is from
  clusters, where the kSZE is relatively unimportant.}
\label{all_hist}  
\end{figure}

We first investigate the contribution of
the WHIM gas to the SZE signal in a sky survey. 
The image histograms for the LDW SZE projections are shown in Figure
\ref{whim_hist}. These histograms are from the raw light cone images,
which have a native angular resolution of 8.8$\arcsec$. It is clear that the
addition of the kSZE creates a strong increase in the number of
pixels above a given flux decrement. The analogous plot for the full
SZE projections (for all gas in the simulations) is shown in Figure
\ref{all_hist}. In this case, the total SZE is dominated by the tSZE,
and the kSZE is a small perturbation on that signal. These figures
illustrate the argument in Section 1.1 very nicely, i.e. the kSZE is a
much more important contributor in the lower density, cooler gas than
it is in the hotter cluster gas. 

Next, we add the tSZE and kSZE maps, and smooth the light cone maps with a Gaussian kernel where the
full width half maximum is equal to the beam size at the thermal
SZE decrement maximum for both Planck Surveyor and the South Pole
Telescope. We can add the two effects for these purposes (adding y and
b) since the relative $\Delta T/T$ at the tSZE maximum decrement is
equal to the value of $-y$, and the relative temperature shift due to
the kSZE is independent of frequency. Under the assumption that our
observation would be made at 150 GHz, this argument is valid. 

\begin{figure}
\includegraphics[width=3.5in]{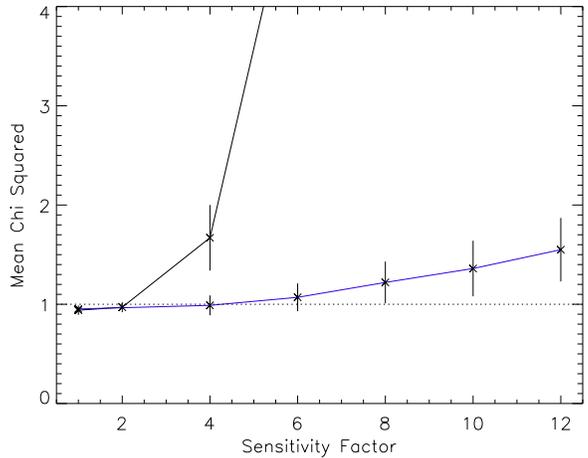}
\caption{The mean value of $\chi^2$ from a Gaussian fit to the image
  histograms for 100 light cone realizations with error 
  bars indicating the range for 90\% of the realizations plotted against the factor by which sensitivity has been
  increased (i.e. noise rms reduced) over the expected value for each
  survey.  A $\chi^2$ value greater than 1 is indicative of a poor fit
  to a Gaussian, and that the filaments contribute signal to the
  maps. Blue is for Planck, black for SPT.}
\label{chi_v_sens}  
\end{figure}

\begin{figure}
\includegraphics[width=3.5in]{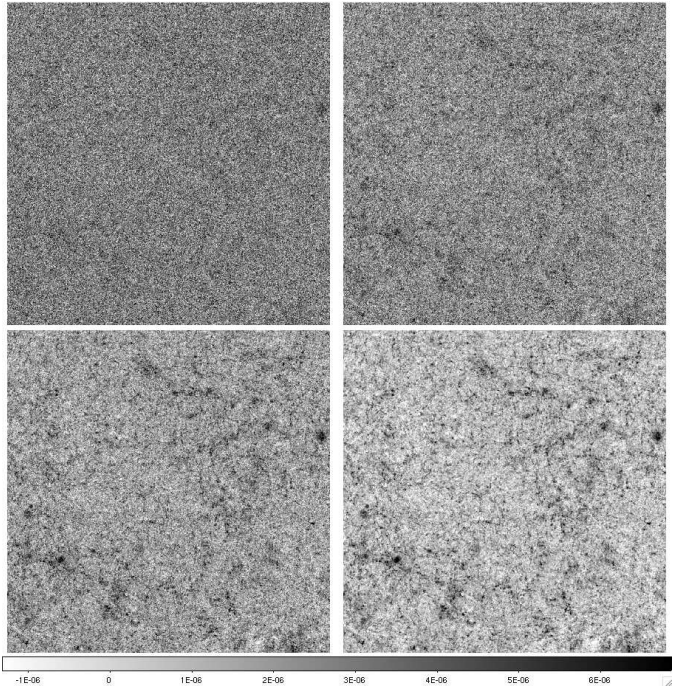}
\caption{Four images of the full 10x10 degree field of the LDW-only
  light cones in total $\Delta T/T$ smoothed with a 1$\arcmin$
  Gaussian to simulate SPT's beam. Upper left: Image
  with 10$\mu$K rms white noise Upper right:image with 5$\mu$K rms,
  lower left: image with 2.5$\mu$K rms, lower right: 1.25$\mu$K rms.}
\label{4panel}
\end{figure}

\begin{figure}
\begin{minipage}[b]{0.96\linewidth}
\includegraphics[width=3.5in]{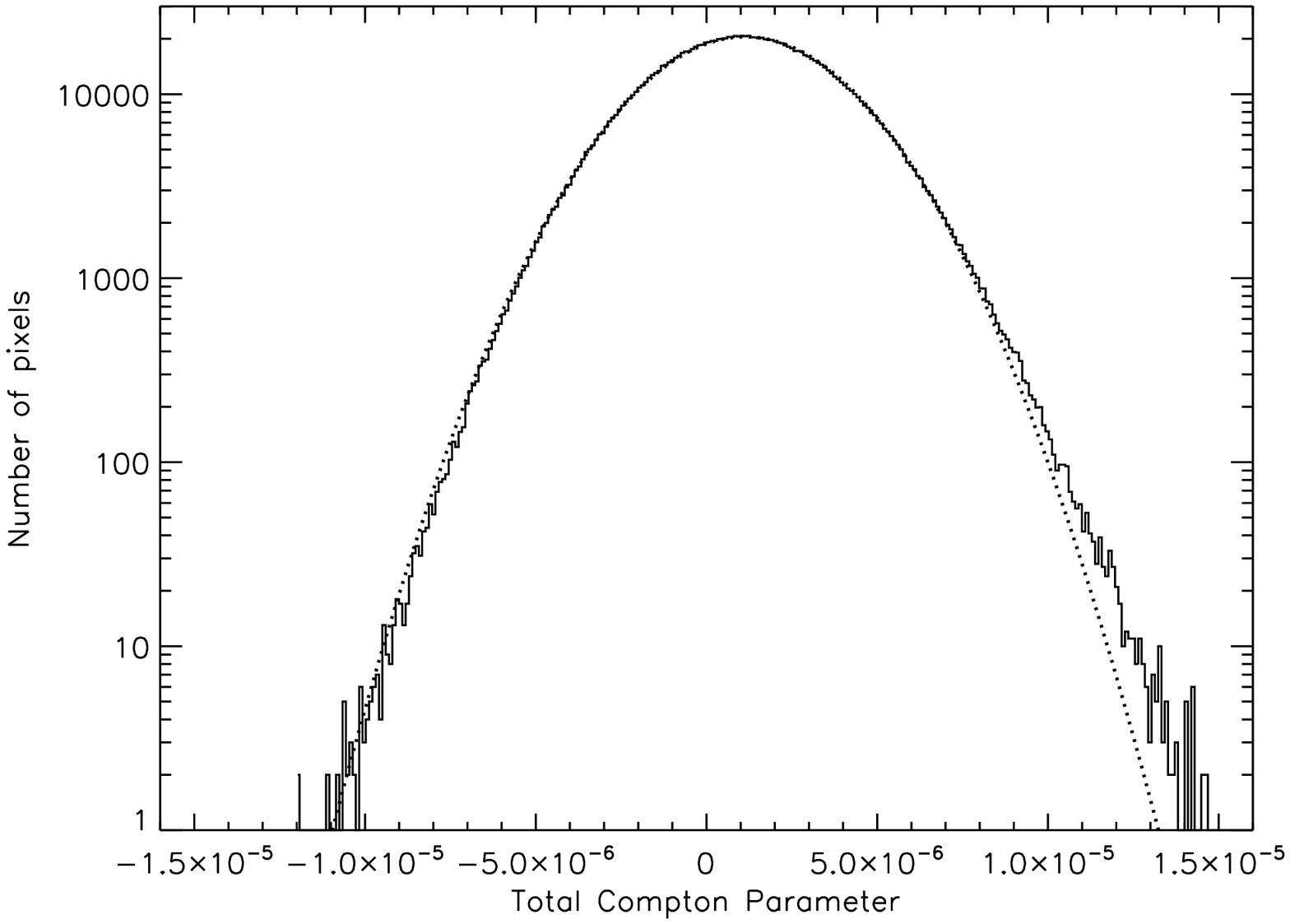}
\caption{Gaussian fit to the image histogram for WHIM light cone with
  SPT beam and a factor of 4 increase in sensitivity over that
  expected in SPT's large survey. This is equivalent to a $\Delta$T
  rms of 2.5$\mu$K per beam.}
\label{fit_spt}
\end{minipage}
\end{figure}

Finally, we add a white noise component with the rms equal
to the expected sensitivity of the two surveys, make a histogram of
image values, and attempt to fit a Gaussian to the histogram.  If the
signal is below the sensitivity of the survey, the fit to a Gaussian should be
excellent.  If the fit is poor, then a
signal has been detected. Figure \ref{chi_v_sens} shows the mean
$\chi^2$ values for the Gaussian fitting for 100 realizations of the
LDW-only light cone. It is clear that for a four-fold increase in
sensitivity in SPT's survey, the LDW makes a real contribution to the
sky signal, and at even a factor of 2 increase (to 5$\mu$K rms per beam), there
might be a marginal detection. Figure \ref{4panel} shows the result of
increasing sensitivity (reducing noise) in the full (10 x 10 degree)
images using the SPT beam size. A sample image histogram for SPT is
shown in Figure \ref{fit_spt}.

\begin{figure}
\begin{minipage}[b]{0.96\linewidth}
\includegraphics[width=3.5in]{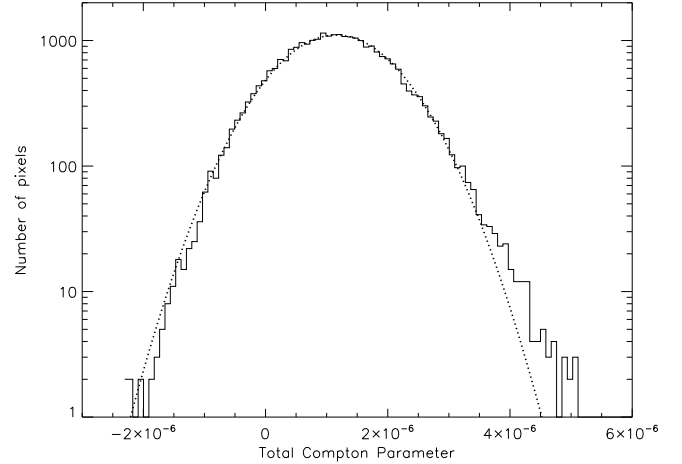}
\caption{Gaussian fit to the image histogram for WHIM light cone with
  Planck beam and a factor of 8 increase in sensitivity over that
  expected in Planck's large survey. This is equivalent to a $\Delta$T
  rms of 0.75 $\mu$K per beam.} 
\label{fit_planck}
\end{minipage}
\end{figure}

For Planck, non-Gaussianity appears
marginally at a factor of 8 increase in sensitivity (0.75 $\mu$K per beam), and is clearly there at a
factor of 10 increase. The difference between these two surveys which
favors SPT is the angular scale of the beam. SPT has about
an arcminute beam at this frequency, while Planck's is
7.1$\arcmin$. The scale of the filaments (of order Mpc),
which is similar to the cluster scale, is a better match to the SPT
beam size for most redshifts. A sample image histogram from an image
smoothed to Planck resolution with its Gaussian fit
overlaid is shown in Figure \ref{fit_planck}. 

We again note that our simulations assume temperature equilibrium between the
ions and electrons, which is probably not achieved on short timescales
in this gas.  This result should reduce the thermal SZE signature by
some amount which is uncertain. A survey would have to be at least as
sensitive as what we have described in order to have WHIM contribution
to the tSZE signal. 

\subsection{Methods for Detecting WHIM Gas in SZE Sky Surveys}
While it appears clear that at a given survey sensitivity the WHIM
gas in filaments (LDW) contributes signal above the noise, it is not
initially obvious how such a measurement would be made.  In the
real universe, we do not have maps of the filaments only, and the lack
of redshift diminution in the SZE results in a significant amount of
confusion in the survey fields. Therefore, separating the contribution
of WHIM from that of clusters and groups of galaxies may seem an
insurmountable challenge. 

\begin{figure}
\includegraphics[width=3.5in]{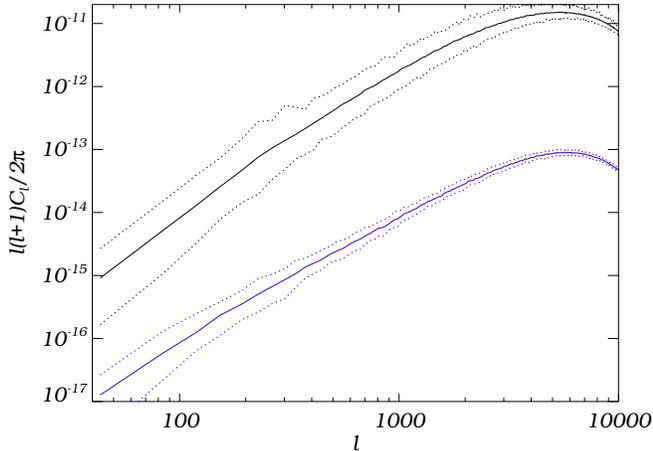}
\caption{Angular power spectra of the thermal SZE from 100 light cone
  realizations with ranges (dotted) within which 90\% of our
  realizations fall. Black is for
  projections of the full simulation, blue is for the low-density WHIM
as described in the text.}
\label{power_th_comp}
\end{figure}

\begin{figure}
\includegraphics[width=3.5in]{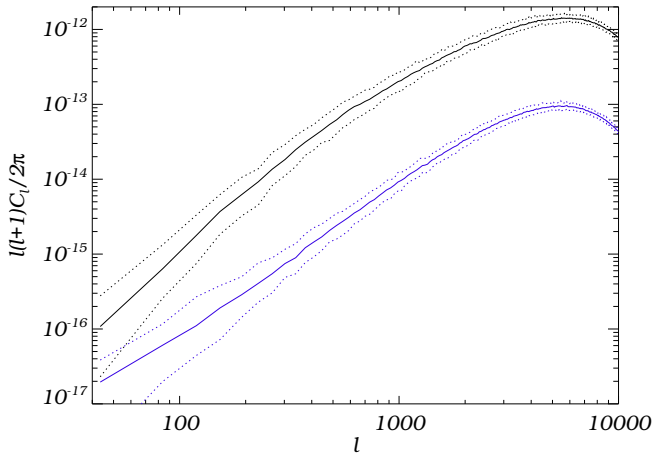}
\caption{Same as Figure \ref{power_th_comp}, but for the kinematic
  SZE. Blue is the angular power spectrum of the low-density WHIM,
  black is full light cone, dotted are for the 90\% range of 100 light
  cone realizations.}
\label{power_k_comp}
\end{figure}

However, it is possible that filaments have a unique angular power signature,
such that their contribution could be identified separate from
clusters. There are significant difficulties with this approach, as illustrated in Figures \ref{power_th_comp} and
\ref{power_k_comp}. These figures show the relative power for a full
simulated sky map and for the low-density WHIM only projections. Figure
\ref{power_th_comp} shows this difference for the tSZE, Figure
\ref{power_k_comp} for the kSZE. While there is a significant amount
of gas in the WHIM phase in simulations, they contribute very little
to the angular power spectrum, particularly in the tSZE. The angular
power at the peak is $\sim$3 orders of magnitude lower in the WHIM for
the tSZE, and more than a full decade lower for the kSZE. More
troublesome is that the shapes of the power spectra are very
similar. We should note that for the kSZE, some small changes in the
stacking algorithm must be made (compared to \citet{sflc}) in order to
maintain continuity of structures along the projection depth. This
means that no shifting in light cone stacking takes place unless the
depth in projection is equal to the depth of the full simulation box. 

Indeed, at these large angular scales the SZE is swamped by the primordial CMB
anisotropies, and the WHIM contributes a very tiny fraction to
the power at these multipoles. What is important to note in these
plots is that the prediction of previous investigators of a bump in the
angular power spectrum of the kSZE at $l \approx 400$ \citep{atrio}
as a result of filaments in the low redshift universe is apparently
absent from our result. This is unfortunate, as this bump would create a unique
angular signature which one could use to identify the WHIM in
filaments. The \citet{atrio} work uses an analytic formalism to
describe the distribution of gas in filaments, and the distribution of
the filaments themselves. It is not clear which assumptions of their
model are not reproduced in our simulations, and could account for the
discrepancy. We will explore this further in later sections. 

\begin{figure}
\includegraphics[width=3.5in]{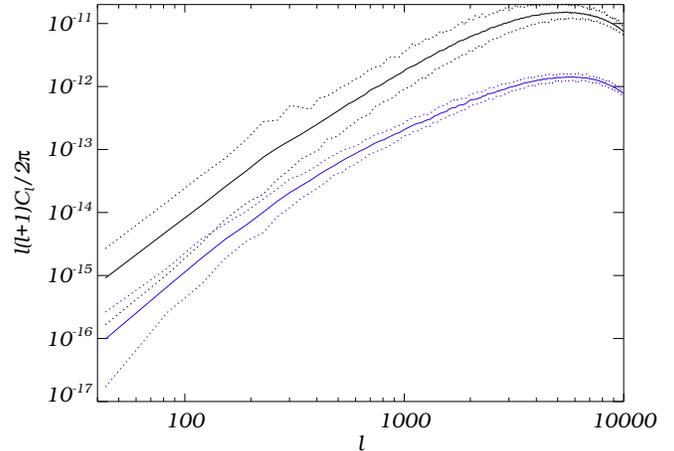}
\caption{Angular power spectra of the full light cone image using all
  gas in the simulation volume. Black is for the tSZE, blue for the
  kSZE, with 90\% variance of 100 light cone realizations indicated by
  dotted lines.} 
\label{power_all_comp_k_t}
\end{figure}

\begin{figure}
\includegraphics[width=3.5in]{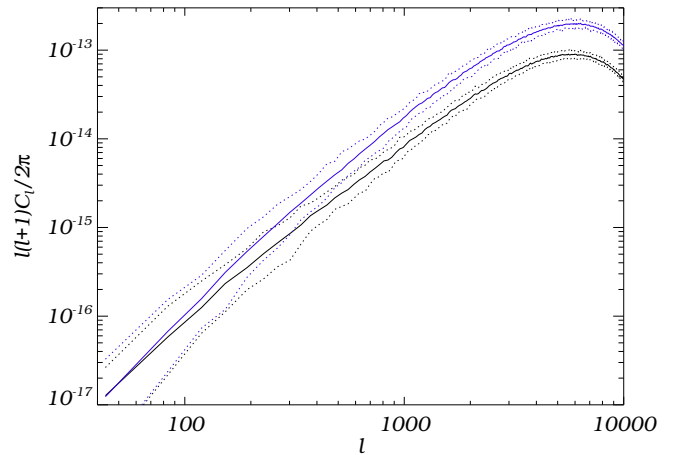}
\caption{Same comparison as Figure \ref{power_all_comp_k_t}, but for
  LDW only light cones. Note that the kSZE has more angular power than
the tSZE for the gas in the LDW phase.}
\label{power_whim_comp_k_t}
\end{figure}

It is interesting to note, however, that the relative power in the
kSZE and tSZE as a function of gas phase varies dramatically. Figure
\ref{power_all_comp_k_t} shows the power spectra for the full light
cone maps with tSZE in black and kSZE in blue. Figure
\ref{power_whim_comp_k_t} shows the same plot for the LDW
gas.  In this gas, the kSZE dominates at the peak and at most
scales. This result meshes nicely with the description of the
kSZE/tSZE ratio in Section 1.1.

\subsection{Effect of the Low-Density WHIM on the Power Spectrum}
Aside from the obvious question of whether the filamentary LDW can be directly detected in
SZE surveys, there is the related key issue of the WHIM filaments
creating a contaminating signal that might be misinterpreted. For
instance, shifts in the SZE angular power spectrum might result in
shifts in fitted cosmological parameters (e.g., the value of
$\sigma_8$). Therefore, it is important to understand the WHIM
contribution to the power spectra. 

To investigate this question, we have taken a ratio of our full simulation light
cone projection power spectra to power spectra of images where the
projection of the LDW SZE has been removed. In all cases we have
smoothed the image with a 1$\arcmin$ Gaussian to represent the beam of
SPT at 150 GHz. We have again done this
for 100 light cone realizations, and plotted the 90\% cosmic variance scatter in
the following figures. Figure \ref{thermal_spt_ratio} shows this ratio
as a function of multipole number for the thermal SZE. While there is
significant cosmic variance at large scales, the mean value (solid
line) is a change of order 3\% maximum. At
high $l$, the ratio is of order 1\%. For the
kSZE, the effect is more important. The lines on Figure
\ref{kinetic_spt_ratio} show the mean
and 90\% scatter of the power spectra ratios. The removal of the
WHIM lowers the overall power in the map by as much as
$\approx $8\% at $l \approx 100-400$. Though declining in contribution
to higher $l$, at $l = 1000$ the effect is still roughly 5\%. 

\subsection{WHIM Power in the kSZE}
While this effect is not obvious when looking
at the WHIM and total power spectra of the images separately, the range of $l$
where the LDW contributes to the power spectrum is broadly consistent with the
results of \citet{atrio}. Though the cosmic variance using different
light cones is quite large, with a large sky area one should see the
effect. In
\citet{atrio}, they find a filamentary contribution which moves to
lower $l$ as $z$ decreases, as is expected due to the reduced angular
diameter distance. Our lowest $z$ defined in these light cones is
0.05, where the filamentary contribution in \citet{atrio} appears to
peak between 1500 $< l <$2000. We see the change at lower $l$, where
they see the contribution coming from $z < 0.05$. We
address this issue further below.

\begin{figure}
\includegraphics[width=3.5in]{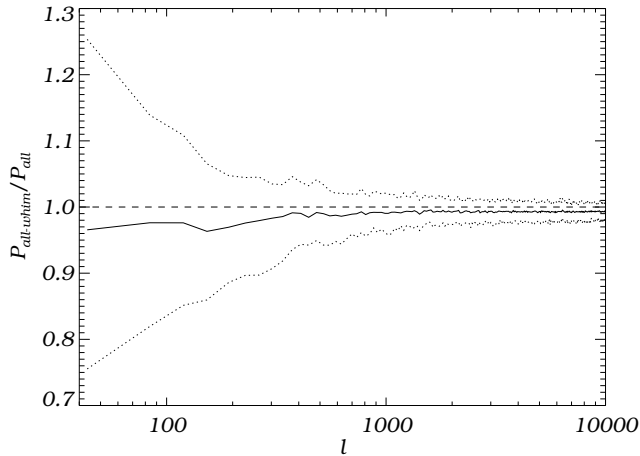}
\caption{Ratio of power spectra from the light cones made from the
  full simulation box with LDW gas subtracted to those made without
  subtracting the LDW for the thermal
  SZE. Dotted lines indicate the 90\% cosmic variance in our 100 light
cone realizations. While there is large variation at low $l$,
a result of the small volume probed at low $z$, the mean is consistent
with very little change due to the presence of LDW gas.}
\label{thermal_spt_ratio}
\end{figure}

\begin{figure}
\includegraphics[width=3.5in]{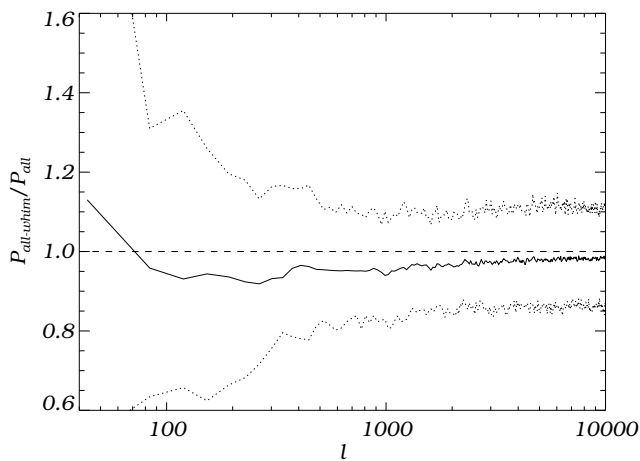}
\caption{Ratio of power spectra as in Figure
  \ref{thermal_spt_ratio}. The removal of the LDW results in a
  reduction of the power at low multipoles. The effect has a maximum
  of $\approx$ 8\% at $l \approx 100-400$.}
\label{kinetic_spt_ratio}
\end{figure}

We predict a filamentary WHIM angular power signature consistent
with \citet{atrio}. There are some differences in the two types of
analyses which may be important. First, \citet{atrio} integrate the
SZE signature of their filamentary models up to relatively high
overdensity (300-500), which in our simulation is outside the
filamentary regime. In addition, our light cones calculate the SZE
signature using a stack of simulation boxes in which the smallest
redshift we use is $z$=0.05. That limit means that low redshift
filaments, which are largest on the sky, are not represented fully by
our method. In their model, the contribution of filaments is a strong
function of redshift, low $z$ being most important. Also, in contrast
to \citet{atrio} we represent a 10x10 degree patch of sky, while their
prediction is for an all sky signature. 

\begin{figure}
\includegraphics[width=3.5in]{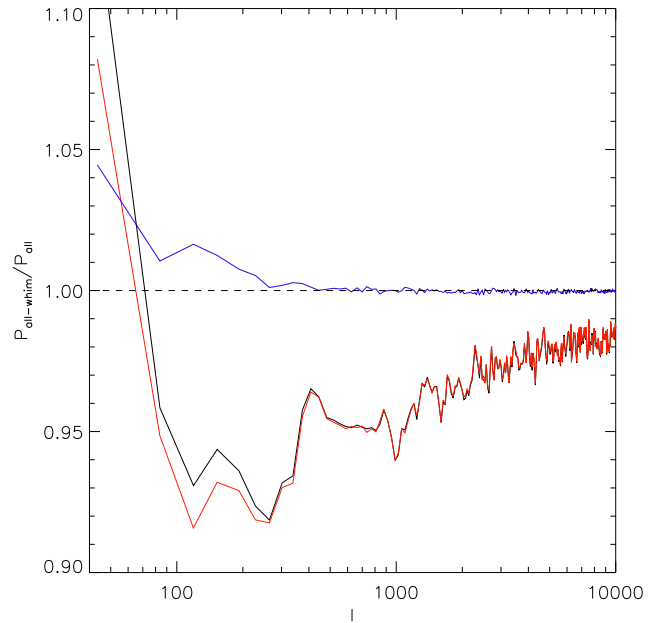}
\caption{Ratio of power spectra as in Figure \ref{kinetic_spt_ratio}
  in black, overlaid with same ratio for light cones adding low z
  slices as described in the text (red line). 5 kSZ slices are added
  from z=0.05 to z=0.0 with a spacing of $\delta$z = 0.01 to
  investigate the power contribution of near field filaments to the
  kSZE power spectrum.  The blue line is the ratio of the two lines,
  with the dotted line at equality for illustration. The low redshift
  filaments do not substantively alter the result. }
\label{near_field}
\end{figure}

To explore the potential contribution of foreground filaments, we have
added to our light cones a set of 5 images spanning the redshifts from
0.05 to 0.00 with a $\delta$z of 0.01 (simulating the work of
\citet{atrio}). For this we have used a smaller simulation of a cubic
volume of dimension 128$h^{-1}$Mpc with 256$^3$ root grid zones and 5
levels of dynamic refinement with refinement criteria similar to those
used in the original simulation. This additional simulation is
necessary since we have much finer $\delta$z spacing, and can then
simulate the contribution of low redshift filaments to the kSZ
images. For these low redshift slices this volume is more than
sufficient, since very little volume is actually sampled in a fixed
angular area of 100 square degrees at $z < 0.05$. We have generated 5
fully independent realizations of the slices at the lowest $z$ range
described above, and added them to images analyzed in earlier
sections. We then repeat the analysis of taking a ratio of power
spectra with and without the LDW filaments. The result is shown in
Figure \ref{near_field}, and it shows that the addition of low $z$
filaments does not substantively change the results. They simply do
not contribute significant power to the overall sky signature in the
kSZE. 

\section{Discussion}
While it appears clear that the LDW gas in filaments contributes a
real signal to the thermal and kinematic SZE above a certain survey
sensitivity, it is unclear how one can take advantage of this effect
to ``detect'' the WHIM gas. The signal associated with the WHIM in the
SZE is washed out by the dominant contributions of the hotter,
denser gas in clusters and groups, and lies in a wavenumber regime
dominated by primary CMB anisotropies. The unique angular
signature that the filaments contribute may make their detection in the
power spectrum possible. The WHIM filaments also make a
contribution to the kSZE power spectrum that could result in
contamination to studies that look to constrain cosmology by the kSZE.

Additionally, it is possible that higher order moments, such as the
angular bispectrum or trispectrum (spherical transform of the three-
and four-point correlation
functions)\citep[e.g.,][]{kogo06} of the images, may be more promising measurements
of filamentary SZE signatures. If the
combined SZE signal of filaments is non-Gaussian \citep[e.g.,][]{yosh01}, a three or four-point
correlation may show a unique signature of filaments. We leave this
analysis to later work. 

This simulation has only adiabatic physics and shocks; no
non-gravitational physics has been included. However, in this density
and temperature regime, cooling times are typically
quite long compared to the Hubble time. Also, what are usually
considered to be the primary non-gravitational heating mechanisms
(stars, AGN) in
the universe are found preferentially in high density
regions. Therefore, the adiabatic approximation is accurate
for this particular problem. However, if episodes of galaxy outflow
heating have taken place in this low density gas (as may be indicated
by its non-negligible metallicity), then corrections may have to be
made. Additionally, the spatial resolution in the low density gas in
this particular simulation is not very high. Our root grid simulation
zones have spatial extent of 1 Mpc, and gas at between overdensity of
1 and 50 has spatial resolution of between 250 and 500 kpc per
zone. Also important is our mass resolution, which means that low mass
(under 5 $\times$10$^{13} M_{\odot}$) clusters are gravitationally
under-resolved. Therefore, gas which is in filaments in our
simulation may properly belong in collapsed structures like
galaxies. It is not immediately clear which direction these effects
would push the results, since higher spatial resolution of the
filaments would result in higher density, increasing the WHIM SZE
signal, and higher mass resolution should result in less gas in the
WHIM phase overall. We will study these questions in later papers.

\section{Summary}
Low-density WHIM gas in filaments contributes signal to 100+ square degree sky
surveys if the sensitivity is sufficiently high. This contribution
becomes important for an SPT-like telescope survey when the
sensitivity is increased by as little as a factor of 2-4x. In the
filaments, the kinematic SZE becomes quite important relative to the
thermal SZE, particularly in the frequency regime of maximum thermal
SZE decrement (150 GHz) where SZE surveys are being performed. In some
phases of the gas, the kSZE can be the dominant contributor to
the SZE signal at that wavelength.  

The presence of WHIM gas in filaments modifies the kinematic SZE
angular power spectrum by $\approx$8\% at $l>100-400$. Our results are
broadly consistent with the prediction of \citet{atrio} of the
$l\approx 400$ bump in the kSZE angular power spectrum due to
filaments. The thermal SZE
is less affected by the presence of filaments. The modification of the
kSZE angular power spectrum could be a source of error in cosmological
parameters derived from the kSZE. It is likely that this bias can be corrected
using numerical simulations like those described in this work. 
The unique angular signature of the WHIM filaments in the kSZE may
allow for WHIM detection using kSZE surveys provided systematic
effects can be accounted for with high accuracy, and a clever method
of isolating the effect from that due to clusters can be devised.
\acknowledgments
BWO and MLN have been supported in part by NASA
grant NAG5-12140 and NSF grant AST-0307690. 
BWO has been funded in part
under the auspices of the U.S.\ Dept.\ of Energy, and supported by its
contract W-7405-ENG-36 to Los Alamos National Laboratory. BWO is supported in part by the NASA ATFP program under grant NNX09AD80G. The
simulations were by performed at SDSC and NCSA with computing time provided by 
NRAC allocation MCA98N020. This work was partially supported by the National Institute for Computational Sciences (NICS) under TG-AST090040 and utilized the NICS Kraken system. EJH and JOB have been supported in part by a grant 
from the U.S. National Science Foundation (AST-0407368). JOB acknowledges support from NSF AST-0807215. EJH also
acknowledges support from NSF AAPF AST-0702923. BDS was supported by
NASA Theory grant NNX07AG77G. Projections were made using the YT analysis toolkit \cite[\texttt{yt.enzotools.org}]{SciPyProceedings_46}. EJH acknowledges useful
discussions with Nils Halverson, Sam Skillman, Licia Verde and Liliya
Williams. We also acknowledge the insightful comments of an anonymous
referee in improving the manuscript.

\bibliographystyle{apj}

\end{document}